\documentclass[11pt]{article}
\pdfoutput=1
\usepackage[T1]{fontenc}
\usepackage[utf8]{inputenc}
\usepackage{dcolumn}
\usepackage{jcappub}

\newcommand{\rmd}{\mathrm{d}}   
\newcommand{\rmi}{\mathrm{i}}   
\newcommand{\rmv}{\mathrm{v}}   
\newcommand{\Tr}{\mathrm{Tr}}   

\newcommand{\Gf}{G_\text{F}}
\newcommand{\Na}{N_\text{a}}

\newcommand{\vac}{\text{vac}}
\newcommand{\matt}{\text{matt}}
\newcommand{\nunu}{\protect{\nu\nu}}

\newcommand{\sfH}{\mathsf{H}} 
\newcommand{\sfM}{\mathsf{M}} 
\newcommand{\sfL}{\mathsf{L}} 
\newcommand{\sfOmega}{\mathsf{\Omega}}

\begin{document}

\title{Multipole expansion method for supernova neutrino oscillations}

\author{Huaiyu Duan,}
\emailAdd{duan@unm.edu}
\affiliation{Department of Physics \& Astronomy, University of New
  Mexico, Albuquerque, NM 87131, USA} 
\author{Shashank Shalgar}
\emailAdd{shashankshalgar@unm.edu}

\keywords{14.60.Pq, 97.60.Bw, 95.30.Cq, 14.60.Lm}

\date{\today}

\abstract{
We develop a multipole expansion method to calculate collective neutrino
oscillations in supernovae using the neutrino bulb model. We test this
method with two representative neutrino energy
spectra, and we find that
this method can be much more efficient than the angle-bin method previously
used in the literature.
The multipole expansion method also provides interesting
insights into multi-angle calculations that were accomplished
previously by the angle-bin method.}

\maketitle
\flushbottom

\section{Introduction}
It is well known that neutrino oscillation probabilities can be
modified by ambient matter through the Mikheyev-Smirnov-Wolfenstein
(MSW) effect \cite{wolfenstein1978neutrino,Mikheev:1986gs}. Likewise,
neutrino oscillations can also be affected by the presence of ambient
neutrinos \cite{Fuller:1987aa,Notzold:1987ik}.
The phenomenon of collective neutrino
oscillations (e.g., \cite{Kostelecky:1993yt, Duan:2005cp,
  Duan:2006an, Hannestad:2006nj, Raffelt:2007cb, Duan:2008za}; see
\cite{Duan:2010bg} for a recent but incomplete 
review), caused by neutrino self-interaction via the 
$Z$-boson mediation, continues to surprise us with new features and
instabilities (e.g., \cite{EstebanPretel:2008ni, Dasgupta:2009mg,
  Gava:2009pj, Friedland:2010sc, Duan:2010bf, Cherry:2012zw,
  Cherry:2013mv, Raffelt:2013rqa,deGouvea:2012hg, deGouvea:2013zp}).

Unlike the MSW effect, collective neutrino flavor transformation caused by
neutrino self-interaction couples the quantum states of neutrinos
themselves and, 
therefore, is nonlinear. Except for a few very simplistic models (see,
e.g., \cite{Kostelecky:1994dt,Hannestad:2006nj,Duan:2007mv}), the
flavor evolution of a dense neutrino gas can be solved only through
numerical methods. This non-linearity together with the inhomogeneous
and anisotropic physical environment
makes it a daunting task to solve collective neutrino oscillations in a
(core-collapse) supernova even numerically. The most sophisticated
calculations of 
this kind so far have  adopted the ``neutrino bulb model'' \cite{Duan:2006an}
which assumes spherical symmetry for the supernova environment.
The current approach of solving neutrino oscillations in the neutrino bulb model
is to discretize the neutrino emission (zenith) angle as well as the
neutrino energy and solve millions to tens of millions of coupled
differential nonlinear equations simultaneously. Although this
``angle-bin method'' is straightforward to implement,
a severe drawback of this approach is that
a large number of angle bins
($\gtrsim 1000$) are required to achieve 
numerical convergence even in the 
regime where no significant neutrino oscillation is observed
\cite{Duan:2006an,Sarikas:2012ad}.

Because of its spherical symmetry, the neutrino bulb model does not
completely capture the complexity of the 
supernova environment revealed in recent multi-dimensional supernova
simulations (e.g., \cite{Bruenn:2012mj, Tamborra:2013laa,
  Dolence:2014rwa}). Even if the supernova is approximately
spherically symmetric, the spherical symmetry in neutrino
oscillations can be broken spontaneously because of the vector-vector
coupling nature of the neutrino self-interaction
\cite{Raffelt:2013rqa, Mirizzi:2013rla, 
  Duan:2013kba}. In the cases where the spherical symmetry is indeed
maintained, the current technique is still insufficient for solving 
neutrino oscillations in the presence of the neutrino halo
\cite{Cherry:2012zw} 
and/or transition magnetic moments of Majorana neutrinos
\cite{deGouvea:2012hg,deGouvea:2013zp}.
It is clear that a new computing model or models are necessary to
address the above challenges. Because the new model(s) will be more
complicated and computationally more demanding than the neutrino bulb
model, it is also clear that the angle-bin method needs to
be replaced by a more efficient approach.

A multipole expansion approach using Legendre polynomials was
employed in an earlier study with the focus on the ``kinematical
decoherence'' of a homogeneous neutrino gas \cite{Raffelt:2007yz}. A
different moment expansion method was later proposed for the neutrino
bulb model \cite{Liao:2009ic}.

In this paper we develop a multipole expansion method similar to that
in \cite{Raffelt:2007yz} for the
neutrino bulb model. The rest of the paper is organized as follows. In
Section~\ref{sec:formalism} we explain the formalism of the new
multipole expansion method. In Section~\ref{sec:discussion} we
test the multipole expansion method with two representative neutrino
energy spectra. 
We also discuss the efficiency of this method, some physical insights
behind it, and how it may be improved.
In Section~\ref{sec:conclusion} we give our
conclusions. Some of the approximations used in our approach are listed
in Appendix~\ref{sec:approximation}.

\section{Formalism%
\label{sec:formalism}}

Here we consider two-flavor neutrino oscillations
(i.e.\ $\nu_e\leftrightarrow\nu_x$ and
$\bar\nu_e\leftrightarrow\bar\nu_x$) in the neutrino bulb model.
Generalization to three neutrino flavors is
straightforward. The neutrino bulb model assumes spherical symmetry
about the center point of the supernova and azimuthal symmetry about any
radial direction from this center point. In this model the
flavor density matrix $\rho_{E,u}(r)$ of the neutrino at radius $r$
depends only on the 
energy $E$ and trajectory $u=\cos(2\vartheta_0)$ of the neutrino, where
the (zenith) emission angle $\vartheta_{0}$ is defined with respect to the
normal of the neutrino sphere at radius $R$. We adopt the
convention of the neutrino flavor isospin \cite{Duan:2005cp} so that
the diagonal elements of $\rho_{E, u}$ (in flavor basis) with $E<0$ are
proportional to the negative number densities of $\bar\nu_e$ and
$\bar\nu_x$ with energy $|E|$. 
The flavor density matrices of neutrinos and anti-neutrinos are
normalized in the following manner: 
\begin{align}
\Tr\int_0^\infty \rho_{E,u}\,\rmd E &= 1, &
\Tr\int_{-\infty}^0\rho_{E,u}\,\rmd E 
&= -\frac{\Phi_{\bar\nu}}{\Phi_\nu},
\end{align}
where $\Phi_{\nu}$ and $\Phi_{\bar{\nu}}$ are the total number fluxes of the
neutrino and anti-neutrino, respectively.

The equation of motion for density matrix  $\rho_{E,u}$ is
\begin{eqnarray}
\rmi v_u \partial_r\rho_{E,u}
= [\sfH_{E,u},\, \rho_{E,u}],
\label{eq:eom-angle}
\label{rho:evol}
\end{eqnarray}
where $\partial_r$ is differentiation with respect to radius $r$,
\begin{align}
v_u =\sqrt{1-\left(\frac{R}{r}\right)^2\sin^2\vartheta_0}
= \sqrt{1-\left(\frac{R}{r}\right)^2\left(\frac{1-u}{2}\right)}
\end{align}
is the radial component of the neutrino velocity, and
\begin{align}
\sfH_{E,u} &= \sfH_\vac+\sfH_\matt+\sfH_\nunu
\nonumber\\
&=\frac{\sfM^2}{2E} + \sqrt2 \Gf \sfL
+\frac{\sqrt2\Gf \Phi_\nu}{2\pi R^2} 
\int_{\sqrt{1-(R/r)^2}}^1\rmd v_{u'}\int_{-\infty}^\infty\rmd E'\, 
(1-v_u v_{u'})\rho_{E',u'}
\label{eq:H}
\end{align}
is the Hamiltonian. In the above equation
$\sfM^2$ is the neutrino mass-squared matrix in flavor
basis, $\Gf$ is the Fermi coupling constant, and $\sfL$ is the matrix of net
charged-lepton number densities. 
Here, for simplicity, we have assumed that neutrinos are emitted isotropically
from the neutrino sphere.  

We define the $n$'th multipole/moment of the neutrino flavor matrix to be
\begin{align}
\rho_{E,n} = \int_{-1}^1 \rho_{E,u} P_n(u)\,\rmd u,
\end{align}
where $P_n(u)$ are the standard Legendre polynomials.
The reason that we use $u=\cos(2\theta_0)$ instead of $v_u$ (as in
\cite{Raffelt:2007yz,Liao:2009ic}) in the definition of the multipoles
is that $u$ has a fixed range $[-1,1]$ but the range of $v_u$ is
$[1-\sqrt{1-(R/r)^2}, 1]$ which changes with $r$.
Using the normalization condition
\begin{align}
\int_{-1}^1 P_m(u) P_n(u)\,\rmd u = \frac{2}{2n+1}\,\delta_{m n}
\label{ortho:con}
\end{align}
we obtain
\begin{align}
\rho_{E,u} = \sum_{n=0}^\infty \left(n+\frac{1}{2}\right) \rho_{E,n} P_n(u).
\label{rho:series}
\end{align}

Instead of transforming eq.~\eqref{eq:eom-angle} to multipole basis directly,
we note that (multi-angle) neutrino oscillations in the neutrino bulb model
usually occur at 
\begin{align}
z \equiv \frac{R^2}{4r^2}\ll1.
\end{align}
Therefore, we first expand eq.~\eqref{eq:eom-angle} in terms of $z$ as in
\cite{EstebanPretel:2008ni,Banerjee:2011fj}.
Using the appropriate lowest order approximation for each term in 
eq.~\eqref{eq:H} we obtain  (see Appendix \ref{sec:approximation})  
\begin{align}
\rmi \partial_r \rho_{E,u} \approx 
[\sfOmega -u \lambda \sigma_3+ \mu (2-u)\rho_0 - \mu \rho_1,\rho_{E,u}],
\label{eq:approx}
\end{align}
where 
\begin{align}
\sfOmega &=\frac{\omega}{2}
\begin{bmatrix} 
-\cos2\theta_{\mathrm{eff}} & \sin2\theta_{\mathrm{eff}}
\\ \sin2\theta_{\mathrm{eff}}  & \cos2\theta_{\mathrm{eff}} 
\end{bmatrix},
\label{H:vac}\\
\lambda(r) &= \frac{\Gf n_e(r) z(r)}{\sqrt2}
\\
\intertext{with $n_e$ being the net electron number density,
  $\sigma_{3}$ is the third Pauli matrix,} 
\mu(r) &= \frac{\sqrt2\Gf \Phi_\nu z^2(r)}{2\pi R^2},\\
\intertext{and}
\rho_n(r) &= \int_{-\infty}^\infty \rho_{E,n}(r)\,\rmd E.
\end{align}
In eq.~\eqref{H:vac}, $\omega \equiv \Delta m^{2}/2E$ is the vacuum
oscillation frequency of the neutrino with $\Delta m^{2}$ being the
mass-squared difference of neutrino, and $\theta_{\mathrm{eff}}\ll1$ is
the effective mixing angle of neutrino in matter
\cite{Duan:2005cp,Hannestad:2006nj}. 
 
Using  identity
\begin{align}
\int_{-1}^1 P_m(u) P_n(u) u\,\rmd u
= \frac{2(n+1)}{(2n+1)(2n+3)}\,\delta_{n+1,m}
+ \frac{2n}{(2n-1)(2n+1)}\,\delta_{n-1,m}
\end{align}
we rewrite eq.~\eqref{eq:approx} in the multipole basis,
\begin{align}
\rmi \partial_r \rho_{E,n} \approx [\sfOmega+\mu(2\rho_0-\rho_1), \rho_{E,n}]
- [\lambda\sigma_{3}+\mu\rho_0, a_n\rho_{E,n+1}+b_n\rho_{E,n-1}],
\label{mom:eom}
\end{align}
where
\begin{align}
a_n &= \frac{n+1}{2n+1}, &
b_n &= \frac{n}{2n+1}.
\end{align}

\section{Validation and discussion%
\label{sec:discussion}}
To validate and show the usefulness of the multipole expansion method
we solve eq.~\eqref{mom:eom} numerically (using the multipole expansion method) with two sets of initial neutrino
energy spectra. 
We assume that the matter density is not large enough to suppress collective
neutrino oscillations, and we take $\lambda=0$. For
comparison we also solve eq.~\eqref{eq:eom-angle} numerically (using the angle-bin method) with replacement
\[ \sfH_\vac+\sfH_\matt \longrightarrow \sfOmega.\]
In both kinds of calculations we use mass-squared difference $\Delta
m^{2} = -3 \times 10^{-3}\,\mathrm{eV}^2$ (i.e.\ with the inverted
neutrino mass hierarchy), 
effective mixing angle $\theta_{\mathrm{eff}}=0.01$,  
100 energy bins per neutrino flavor, and the radius of
the neutrino sphere $R=11$ km.  

The two sets of neutrino spectra used in our calculations are adapted from
\cite{Duan:2006an} and \cite{Dasgupta:2009mg}, respectively, which are known to
produce a single swap/split (SS) and multiple swaps/splits (MS) in
final neutrino fluxes. 
In both sets the neutrino spectra
are described by the Fermi-Dirac distribution
\begin{equation}
f_{\nu}(E) \propto
\frac{E^{2}}{1+\exp\left(\displaystyle\frac{E}{T_{\nu}}-\eta_{\nu}\right)},
\qquad 
(\nu=\nu_{e}, \bar{\nu}_{e}, \nu_{x},\bar{\nu}_{x}) 
\end{equation}
with the parameters listed in table~\ref{ini:spec}.
\begin{table}
\begin{center}
\begin{tabular}{l| c c }
\hline\hline
 & SS spectra & MS spectra \cr
 \hline
$L_{\nu_{e}}$ ($10^{51}$ ergs/sec) & $1.0$ & $4.1$\cr
$L_{\bar{\nu}_{e}}$ ($10^{51}$ ergs/sec) & $1.0$ & $4.3$\cr
$L_{\nu_{x}/\bar\nu_x}$ ($10^{51}$ ergs/sec) & $1.0$ & $7.9$\cr
$T_{\nu_{e}}$ (MeV) & 2.8 & 2.1\cr
$T_{\bar{\nu}_{e}}$ (MeV) & 4.0 & 3.4\cr
$T_{\nu_{x}/\bar\nu_x}$ (MeV)& 6.3 & 4.4\cr
$\eta_{\nu_{e}}$ & 3.0 & 3.9\cr
$\eta_{\bar{\nu}_{e}}$ & 3.0 & 2.3\cr
$\eta_{\nu_{x}/\bar\nu_x}$ & 3.0 & 2.1\cr
\hline\hline
\end{tabular}
\end{center}
\caption{The two sets of neutrino luminosities and spectral parameters
  used in our calculations. They are
  adapted from \cite{Duan:2006an} and \cite{Dasgupta:2009mg},
  respectively, and are  
  known to produce a single swap/split (SS) and multiple swaps/splits
  (MS) in final neutrino fluxes. }
\label{ini:spec}
\end{table}

\begin{figure}
\includegraphics[width=0.49\textwidth]{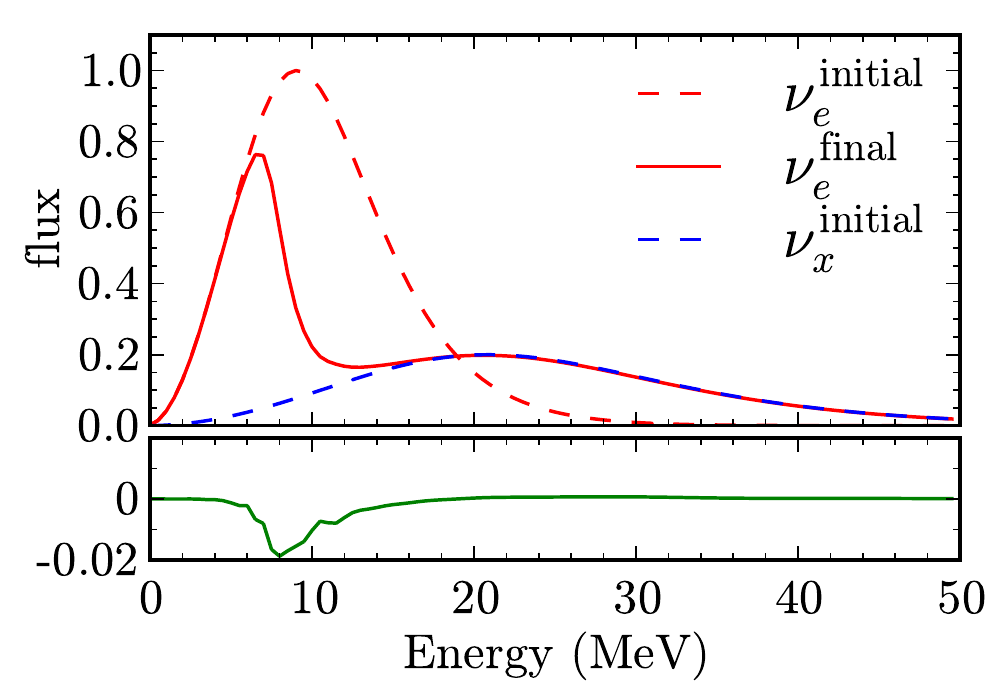}
\includegraphics[width=0.49\textwidth]{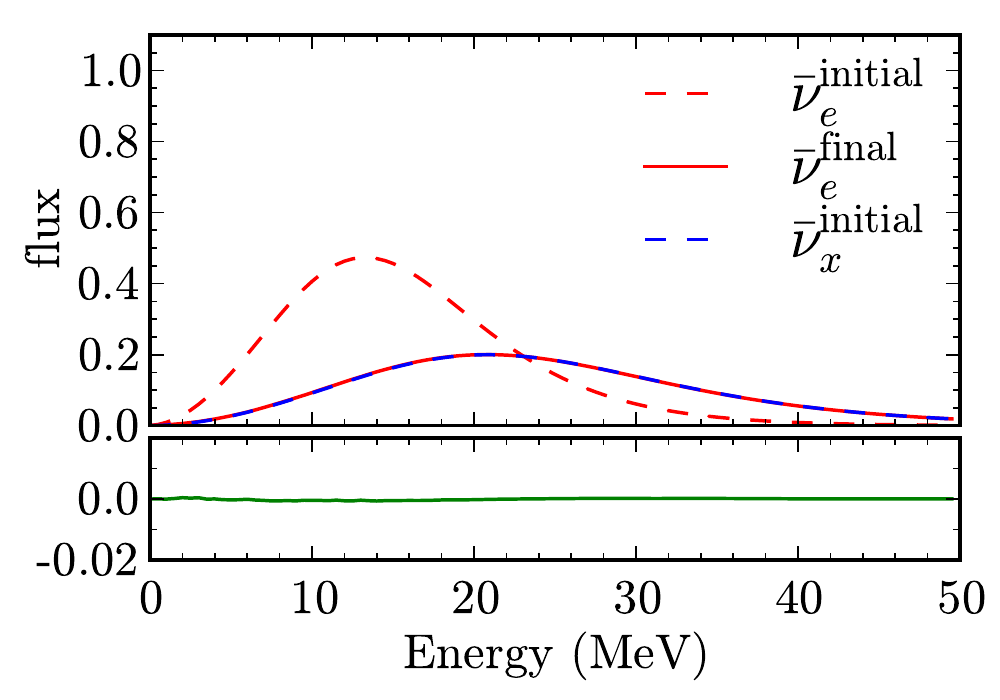}
\caption{Neutrino fluxes (in arbitrary units) in the SS spectrum
  case. The left and right panels are for neutrino and anti-neutrino,
  respectively. In the top panels the dashed curves are for the 
  initial spectra at the
  neutrino sphere, and the solid curves are for the fluxes at 200 km
  computed in multipole basis with 25 multipoles. The bottom
  panels show the differences 
  between the final $\nu_{e}$ fluxes computed in multipole basis and 
  angle basis (with 1200 angle bins). }
\label{ssplit:both}
\end{figure}

In figure~\ref{ssplit:both} we show the neutrino fluxes at 200 km
computed in multipole basis with 25 mutipoles (i.e.\ with
$\rho_{E,n \geq 25} = 0$) and with the SS spectra. 
In the same figure we also plot the differences between these results
and those computed in angle basis with 1200 angle bins.
One can see that these two calculations agree with each other very
well. We note that the small differences between the two results are
partly due to the approximations we made in eq.~\eqref{eq:approx}
which is employed in the multipole expansion method.

To understand why so few multipoles are needed, we consider the
strengths of the multipoles defined as follows, 
\begin{eqnarray}
S_{E,n} &\equiv& \left[\left(2n+1\right)\frac{\Tr \rho^{2}_{E,n}}{(\Tr
    \rho_{E,0})^{2}}\right]^{1/2}. 
\end{eqnarray}
In particular,
$S_{E,0}$ is an indicator of the flavor polarization of the angle
averaged neutrino flux with energy $E$, and $S_{E,0}=1/\sqrt{2}$
implies complete flavor depolarization, i.e.\ equal number of
$\nu_{e}$ ($\bar{\nu}_{e}$) and $\nu_{x}$ ($\bar{\nu}_{x}$).  
Because we consider only the coherent forward scattering of neutrinos
outside the neutrino sphere, $\Tr \rho_{E,u} $ and, therefore,
$\Tr \rho_{E,n}$, do not change with radius. Furthermore, because we do
not consider the quantum decoherence of the neutrino states,
$\Tr \rho_{E,u}^{2}$ are also constant. Using eqs.~\eqref{ortho:con}
and \eqref{rho:series} it is straightforward to show that 
 \begin{equation}
 \frac{\rmd}{\rmd r} \sum_{n} S^{2}_{E,n} = 0.
 \label{sum:con}
\end{equation}

\begin{figure}
\includegraphics[width=0.49\textwidth]{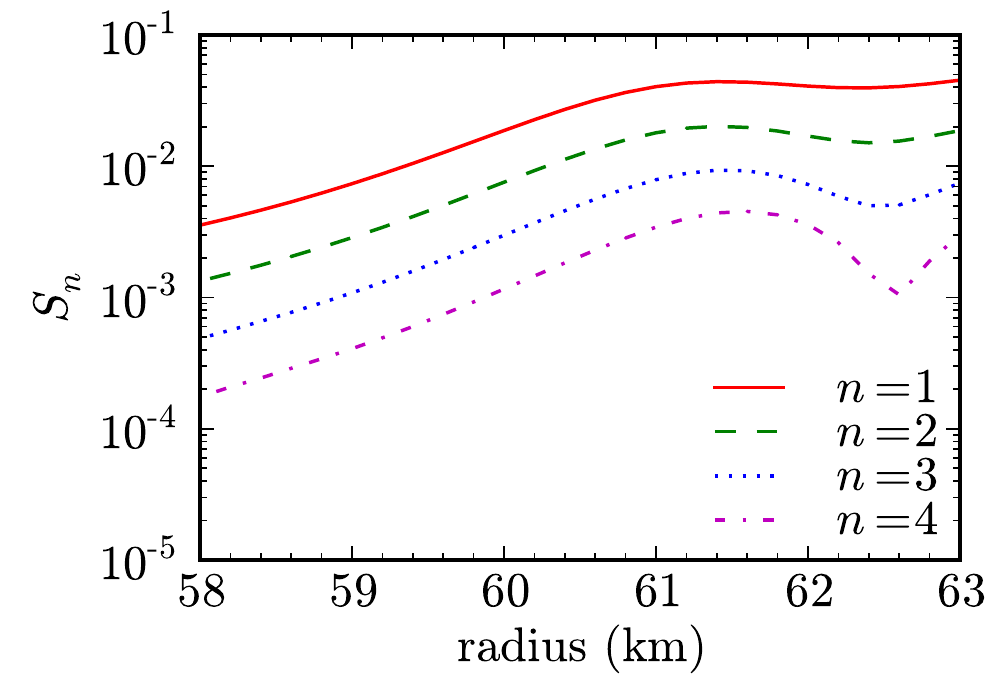}
\includegraphics[width=0.49\textwidth]{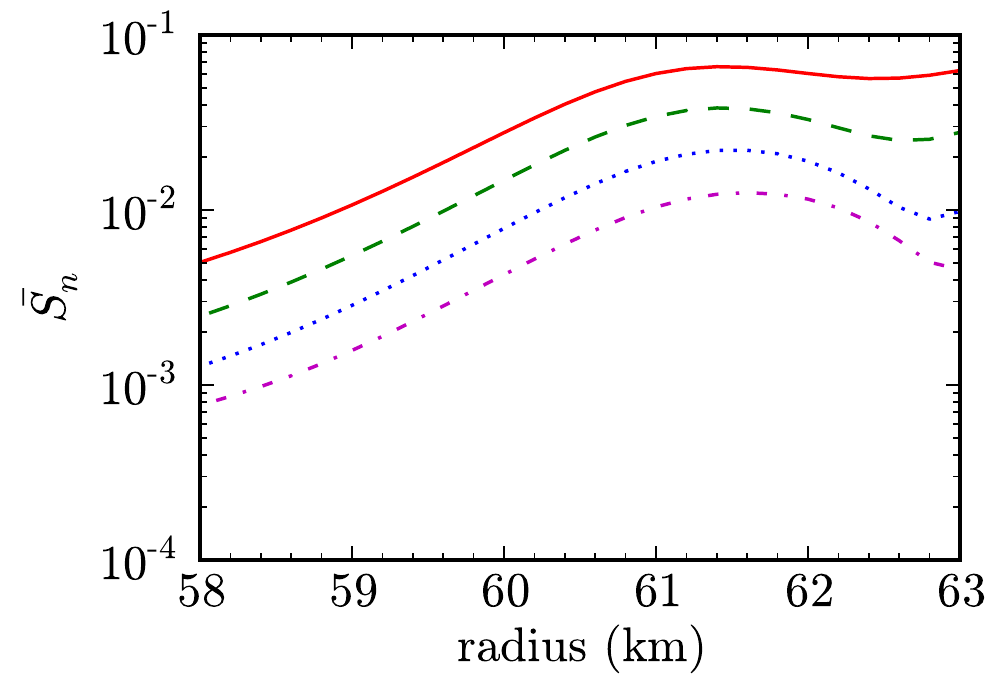}
\includegraphics[width=0.49\textwidth]{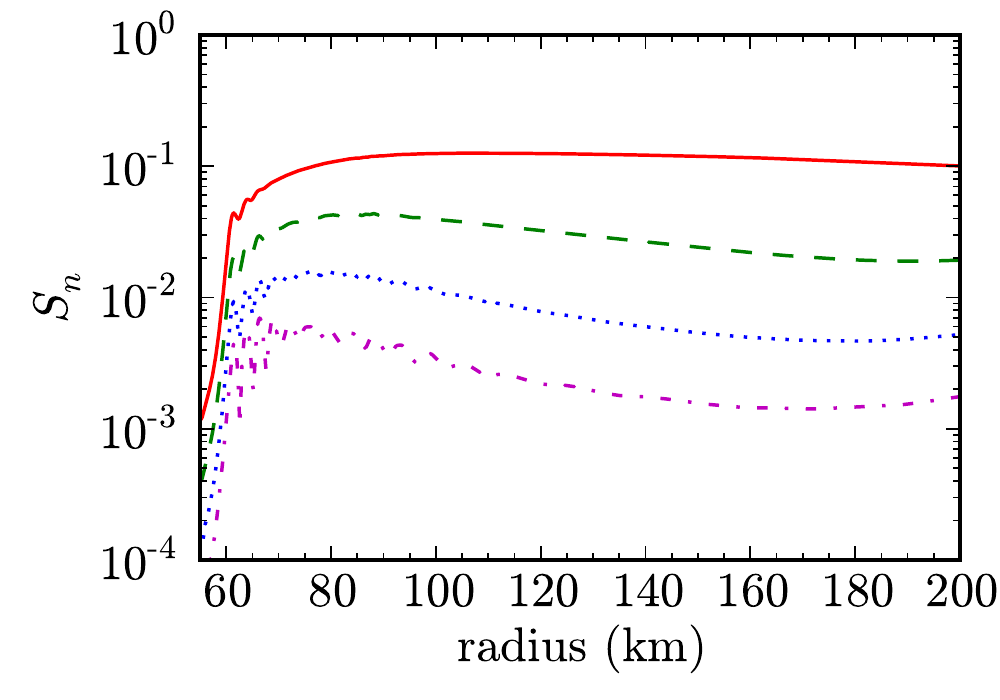}
\includegraphics[width=0.49\textwidth]{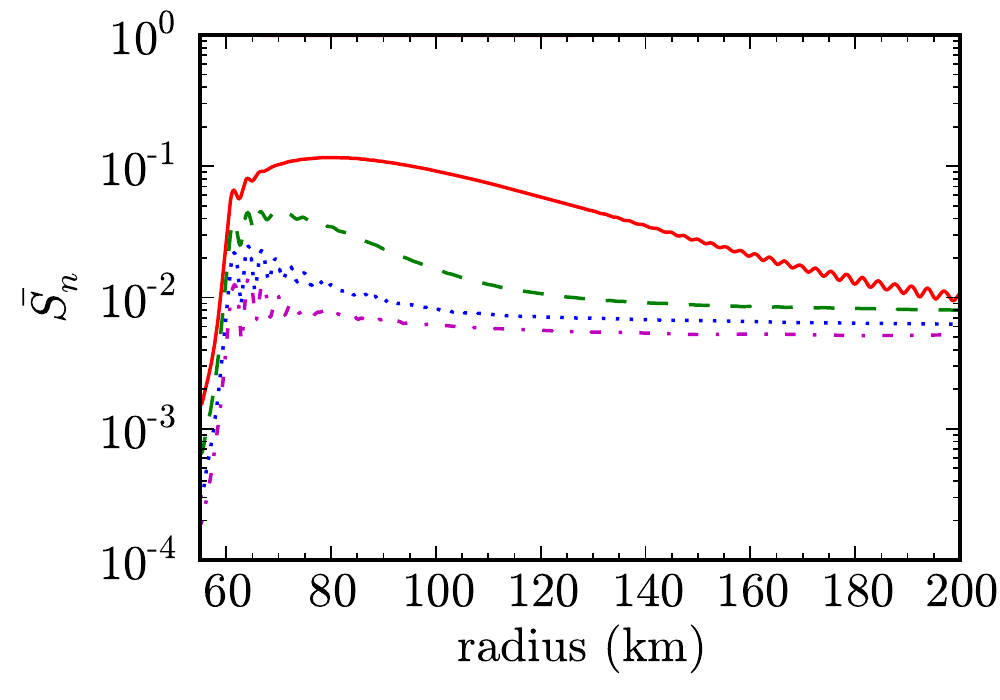}
\caption{The overall strengths of the multipoles of the neutrino fluxes as
  functions of radius in the SS spectrum case. The left and right
  panels are for neutrino and 
  anti-neutrino, respectively. The top panels show the exponential
  growth of $S_{n}$ and $\bar{S}_{n}$ right after collective
  oscillations begin. The bottom panels show the behavior of these multipole
  strengths in the whole regime where collective oscillations are active.}
\label{tr:ssfull}
\end{figure}

In figure~\ref{tr:ssfull} we show how 
\begin{align}
S_{n} &\equiv \left(\int_{0}^{\infty}
S^{2}_{E,n}\Tr\rho_{E,0}~dE\right)^{1/2}, & \bar{S}_{n} &\equiv
\left(\int_{-\infty}^{0} S^{2}_{E,n}|\Tr\rho_{E,0}| ~dE\right)^{1/2}   
\end{align}
evolve as functions of radius. One can see that, right after
collective neutrino oscillations begin, both $S_{n}$ and $\bar{S}_{n}$
grow exponentially with radius. 
This result suggests that all multipoles become unstable
simultaneously which one may expect from the ansatz of the linear
stability analysis in angle basis \cite{Banerjee:2011fj}. However,
this result seems to be contrary to a previous study of the
homogeneous gas of neutrinos where higher multipoles are populated
successively by diffusion from lower multipoles \cite{Raffelt:2007yz}. 

From figure~\ref{tr:ssfull} one can also see that $S_{n}$ and $\bar{S}_{n}$
are never large for $n>0$. This property together with
eq.~\eqref{sum:con} implies that $S_{E,0}$ is almost constant and that
there is no significant flavor depolarization in this particular case. This
result is, of course, already known from figure~3 in
\cite{Duan:2006jv}, which motivated us to solve neutrino oscillations
in multipole basis in the first place. 
From figure~\ref{tr:ssfull} one may conclude that only the first very
few multipoles, far
fewer than 25, are needed for this calculation. This is
indeed true but only up to a certain radius after which spurious
oscillations of big amplitudes would occur. This phenomenon implies
that some kind of diffusion among multipoles (as suggested in 
\cite{Raffelt:2007yz}) may indeed exist while collective oscillations are
active. Such spurious oscillations may be suppressed by
implementing an appropriate closure scheme (as, e.g., in
\cite{korner1992approximate}). 
We note that any closure that insures the accuracy of
the first multipoles would be sufficient for most physical applications
at $r\gg R$.  

It is interesting to note that, 
in the regime where no significant neutrino
oscillations 
occur, all multipoles with $n>0$ are very small. This is because in
this regime all neutrinos
essentially stay in the same flavor states as they are at the neutrino
sphere. Therefore,
one can set all but the first two multipoles to 0 initially and solve
eq.~\eqref{mom:eom} for $\rho_{E,0}$ and $\rho_{E,1}$ only until
the magnitudes of $\rho_{E,1}$ cross certain threshold which depends
on the desired 
precision.\footnote{In the case where the angle 
  distributions of the neutrino fluxes are not isotropic on the neutrino
  sphere, e.g.\ in \cite{Mirizzi:2012wp}, one needs to solve for the first
  several multipoles before collective neutrino oscillations begin.} (Solving
eq.~\eqref{mom:eom} for $\rho_{E,0}$ only is equivalent a
single angle approximation, which may lead to wrong
results \cite{Duan:2010bf}.) More moments can be added adaptively from
this point on as necessary.  

In contrast, the angle-bin method requires
a large number of angle bins ($\Na\gtrsim1000$) to achieve numerical
convergence even in the regime where no physical oscillations occur
\cite{Duan:2006an}. 
It was shown in \cite{Sarikas:2012ad} that the numerical solutions to
eq.~\eqref{eq:approx} with discrete angle bins have flavor
instabilities that are absent in the continuum limit
$\Na\rightarrow\infty$. These spurious flavor
instabilities can be suppressed by employing a large number of angle
bins. Our calculation shows that such spurious oscillations do not
occur, at least in this particular case, in the numerical 
solution to eq.~\eqref{mom:eom} which is completely equivalent to
eq.~\eqref{eq:approx}.

\begin{figure}
\includegraphics[width=0.49\textwidth]{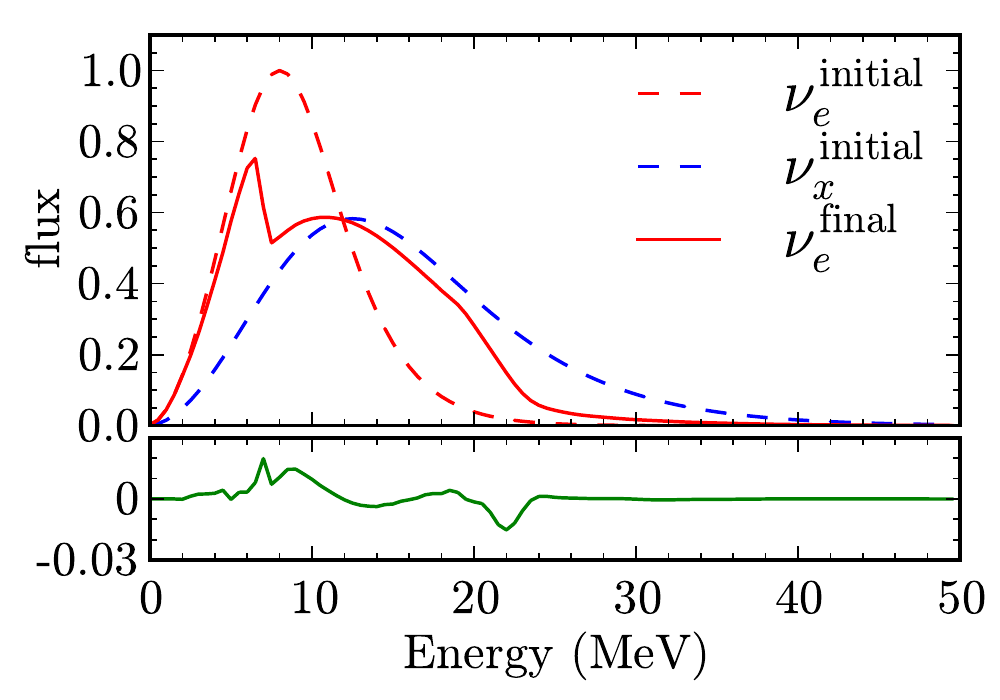}
\includegraphics[width=0.49\textwidth]{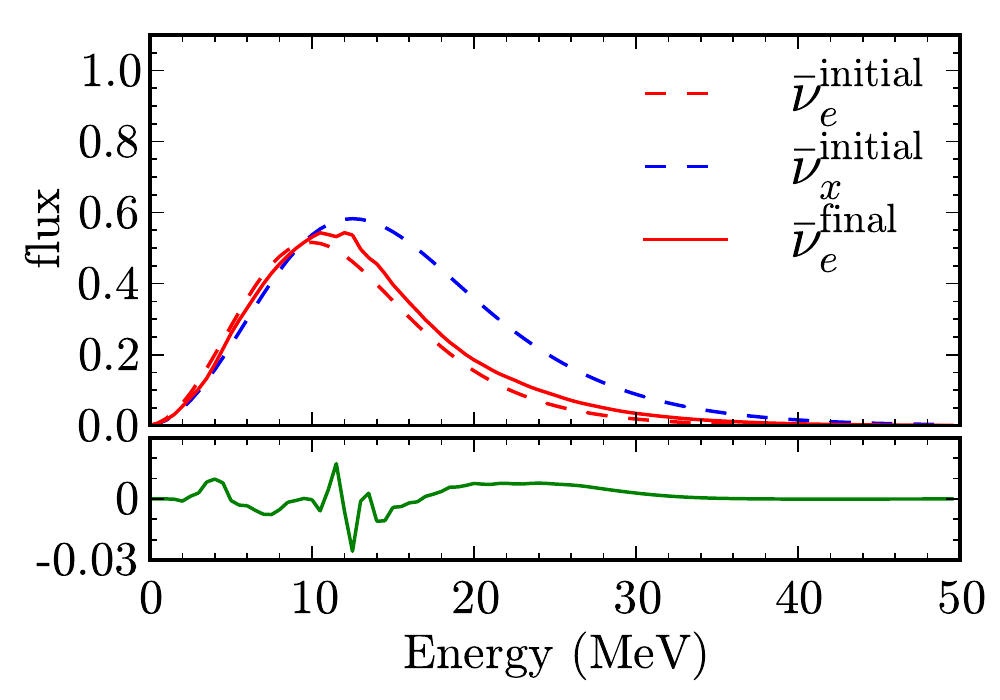}
\caption{Neutrino fluxes at 400 km for the MS spectrum case. The
  notations are the same as in figure~\ref{ssplit:both}. The
  computations was performed with 300 multipoles in multipole basis
  and 50000 angle bins in angle basis.} 
\label{msplit:both}
\end{figure}

\begin{figure}
\includegraphics[width=0.49\textwidth]{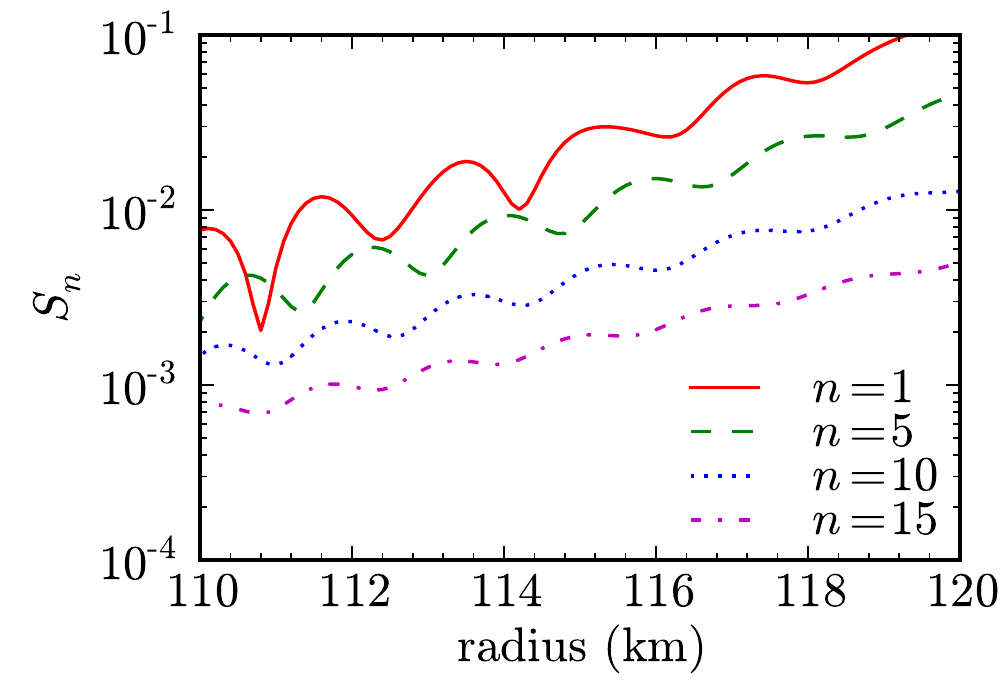}
\includegraphics[width=0.49\textwidth]{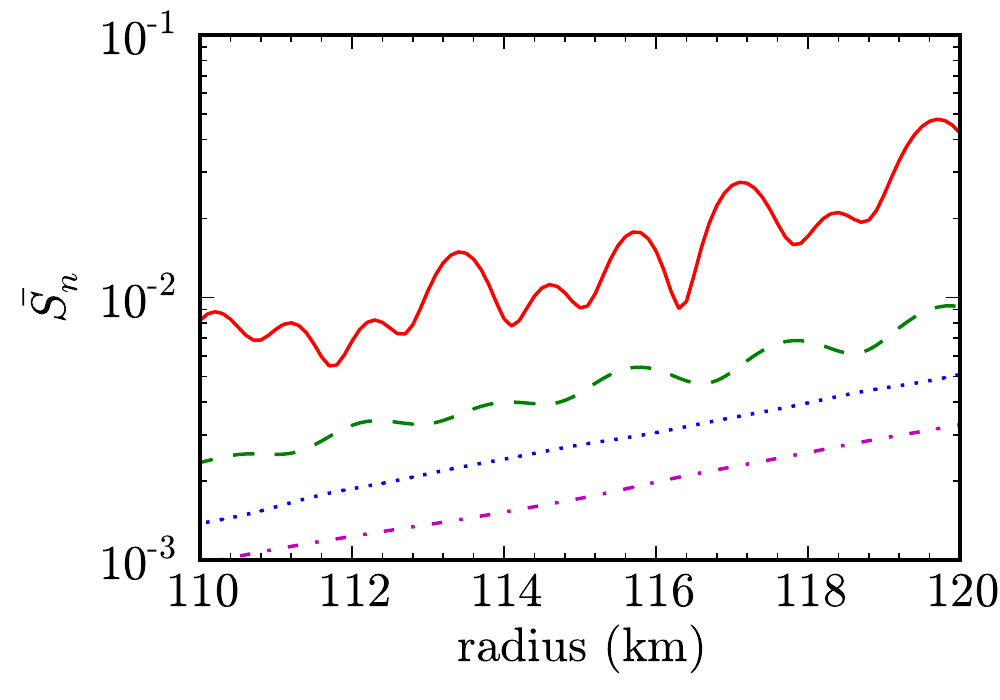}
\includegraphics[width=0.49\textwidth]{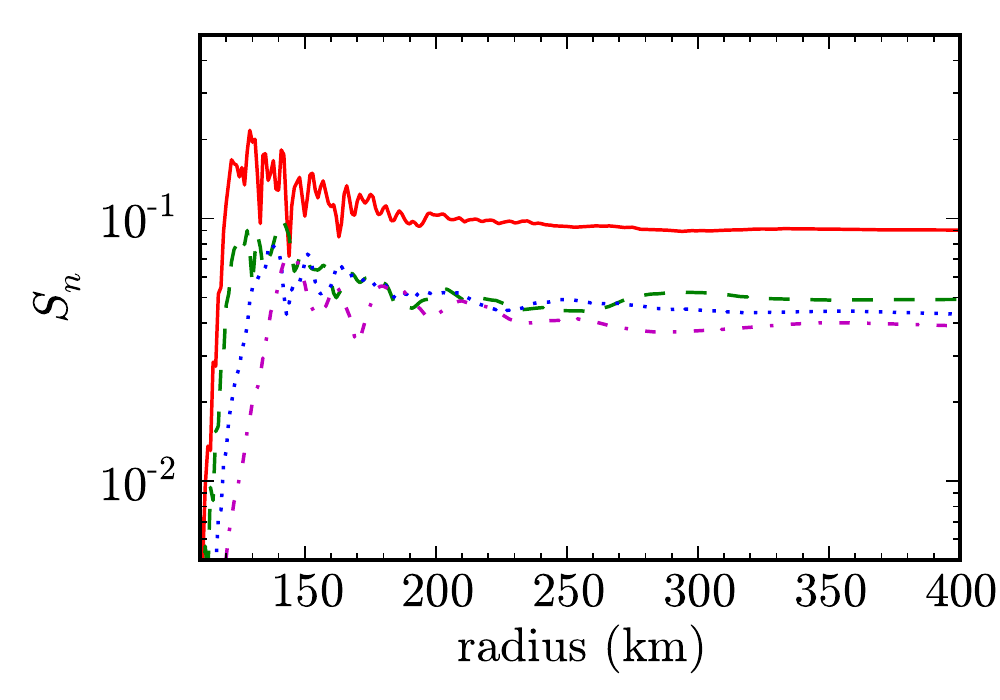}
\includegraphics[width=0.49\textwidth]{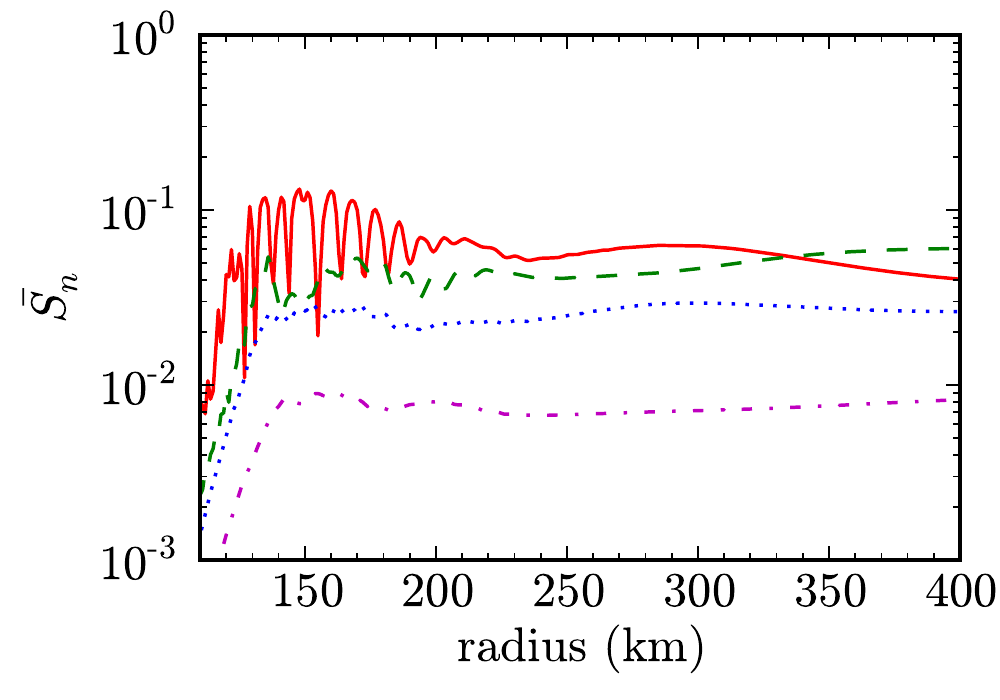}
\caption{Same as figure~\ref{tr:ssfull} but for the MS spectrum
  case.}
\label{tr:msfull}
\end{figure}

We also performed the calculations for the MS spectrum. It was
reported in \cite{Dasgupta:2009mg} that one needed 15000 angle bins to
achieve ``apparent convergence'' for this case. We found that
true convergence could be achieved with 50000 angle bins in our
calculation. In comparison only 300 multipoles 
are required to achieve numerical convergence in  multipole basis. In
figure~\ref{msplit:both} we show the neutrino fluxes at 400 
km in these calculations, and  
in figure~\ref{tr:msfull} we show the multipole strengths $S_n$ and
$\bar{S}_n$ as functions of radius. 

Our discussions for the SS spectrum case also apply to MS spectrum
case. However, we note that the strengths of
the multipoles in this case do not decrease rapidly with increasing
$n$. The first several multipoles oscillate with radius when
collective oscillations begin, and their
strengths are of similar magnitude. 
Significant more multipoles are required to achieve numerical
convergence in the MS spectrum case than
in the SS spectrum case.

\section{Conclusion%
\label{sec:conclusion}}

We have developed a multipole expansion method for the
multi-angle 
calculations of collective neutrino oscillations in supernovae. 
We have tested this method with two representative neutrino energy
spectra. Our calculations show that, at least for these two cases, the
multipole expansion method requires solving far fewer number of equations 
 (by approximately two orders of magnitude) than the traditional
angle-bin method does. 

The relative efficiency of the multipole expansion method are
twofold: (a) Very few multipoles are needed in the regime where no
physical oscillations occur. In contrast, a large number of angle bins
are required to suppress spurious oscillations in the angle-bin
method. (b) The strengths of the multipoles decrease with increasing
multipole index $n$ in the regime where collective neutrino
oscillation are active. Although we have tested the multipole expansion
method for two representative neutrino energy spectra, the
conclusions obtained here may apply to many other situations.
More work is needed to be done in this regard.

The multipole expansion method can potentially be
adapted to more realistic supernova models. The efficiency of this
method may be further improved if an appropriate
closure scheme is employed.

\appendix
\section{Approximations
\label{sec:approximation}}

Like in previous works (e.g., \cite{EstebanPretel:2008ni,Banerjee:2011fj})
we expand the Hamiltonian in eq.~\eqref{eq:H} in terms of $z=R^2/4r^2$
and keep the 
leading order terms for each part of the Hamiltonian. Higher order terms can be
included for more accuracy. 
For vacuum Hamiltonian we have
\begin{align}
\frac{\sfH_\vac}{v_u}
\xrightarrow{z\ll1} 
\frac{\omega}{2}
\begin{bmatrix} 
-\cos2\theta_\rmv & \sin2\theta_\rmv  \\ \sin2\theta_\rmv  & \cos2\theta_\rmv 
\end{bmatrix}
+\mathcal{O}(z),
\end{align}
where $\omega = \Delta m^2/2E$ is the vacuum oscillation frequency
with $\Delta m^2$ being the neutrino mass-squared difference, and
$\theta_\rmv$ is the neutrino vacuum mixing angle. 
We have dropped the trace term which has no impact on neutrino oscillations.
We note that the dispersion of
$\sfH_\vac/v_u$ in neutrino energy $E$, which is much larger than its
dispersion in angle parameter $u$,  
plays an essential role in collective neutrino oscillations
\cite{Duan:2005cp}.  We also
note that the terms of order $z$ or higher are unlikely to be important
because $\sfH_\vac$ is smaller than $\sfH_{\nu\nu}$ where collective
oscillations are active.

For matter Hamiltonian we have
\begin{align}
\frac{\sfH_\matt}{v_u} 
\xrightarrow{z\ll1}
\lambda(r) [z^{-1}+(1-u)]
\begin{bmatrix}
1 & 0 \\ 0 & -1
\end{bmatrix}
+\mathcal{O}(z^2),
\label{H:matt}
\end{align} 
where 
\begin{align}
\lambda(r) = \frac{\Gf n_e(r) z(r)}{\sqrt2},
\end{align}
and where we have again dropped the trace term.
The terms on the right hand side of eq.~\eqref{H:matt} that do not
depend on $u$ 
have no effect on collective neutrino oscillations other than resetting
the vacuum mixing angle $\theta_\rmv$ to a smaller value, which we denote by
$\theta_{\mathrm{eff}}$ 
\cite{Duan:2005cp,Hannestad:2006nj}. 
The angle dependent term in eq.~\eqref{H:matt} can lead to suppression
of collective oscillations in the case of very high matter density,
which can be important near
the neutrino sphere and/or during the accretion phase of the
explosion~\cite{EstebanPretel:2008ni}. The terms of order 
$z^2$ or higher in eq.~\eqref{H:matt} are
unlikely to be important because the matter density itself decreases
rapidly with $r$. 
In the regime where the matter density is much
larger than $\Gf^{-1}|\omega|$ for typical neutrino energies, we obtain
\begin{align}
\frac{1}{v_u}(\sfH_\vac+\sfH_\matt)
\longrightarrow
\frac{\omega}{2}
\begin{bmatrix} 
-\cos2\theta_{\mathrm{eff}} & \sin2\theta_{\mathrm{eff}}  \\ \sin2\theta_{\mathrm{eff}}  & \cos2\theta_{\mathrm{eff}}
\end{bmatrix}
-u \lambda(r) \begin{bmatrix}
1 & 0 \\ 0 & -1
\end{bmatrix}.
\end{align}

For the neutrino self-coupling Hamiltonian we have
\begin{align}
\frac{\sfH_\nunu}{v_u}
\approx
\mu(r) [(2-u)\rho_0 - \rho_1]+\mathcal{O}(z^{3}), 
\end{align}
where
\begin{align}
\mu(r) = \frac{\sqrt2\Gf \Phi_\nu z^2}{2\pi R^2},
\end{align}
and
\begin{align}
\rho_n = \int_{-\infty}^\infty \rho_{E,n}\,\rmd E.
\end{align}
The angle dependence of $\sfH_\nunu$ can lead to the
multi-angle suppression of collective oscillations
\cite{Duan:2010bf,Banerjee:2011fj}.

\acknowledgments{
We would like to thank Sajad Abbar and Vahid Noormofidi for useful discussions.
This work was supported by DOE EPSCoR grant \#DE-SC0008142 at UNM.

}
\bibliography{threef}
\bibliographystyle{JHEP}
\end{document}